%% file: n2_11.tex
\documentclass[twocolumn,preprintnumbers,aps,superscriptaddress]{revtex4}

\usepackage{graphicx}
\usepackage{psfrag}
\usepackage{amsmath}
\usepackage{bm}
 
\input epsf

\input dft

\input psfig.sty
\begin{document}
\title{Undoing static correlation: Long-range charge transfer in time-dependent density functional theory}
\date{\today}
\author{Neepa T. Maitra}
\affiliation{Department of Physics and Astronomy, City University of New York and Hunter College, 695 Park Avenue, New York, NY 10021, USA}
\email{nmaitra@hunter.cuny.edu}

\begin{abstract}
Long-range charge transfer excited states are notoriously badly underestimated
in time-dependent density functional theory (TDDFT).
We resolve how {\it exact} TDDFT captures charge transfer between open-shell species: in particular the role of the step in the
ground-state potential, and the severe frequency-dependence of the
exchange-correlation kernel. An expression for the latter is derived, that becomes exact in the
limit that the charge-transfer excitations are well-separated from
other excitations.  The
exchange-correlation kernel has the task of undoing the static
correlation in the ground state introduced by the step, in order to
accurately recover the physical charge-transfer states.

\end{abstract}
\maketitle 
\section{Introduction}
How or whether excited states of long-range charge transfer character
are captured in time-dependent density functional theory (TDDFT) has
recently received much attention~\cite{DWH03,T03,DH04,GB04b,TAHR99,F01,LLS03,CGGG00,TTYY04}, partly
because of its importance in systems of biological and chemical
interest. Many of these molecules are big enough that traditional
wavefunction methods to calculate excited states become prohibitively
expensive; TDDFT, on the other hand, scales favorably with the number
of electrons,
while remaining reliably accurate for many
excitations. There is therefore much interest in applying TDDFT to
such systems, and understanding why, when, and where it is expected to
work well. 

TDDFT is an exact theory, based on the mapping of the interacting
electronic system to a non-interacting one~\cite{RG84}, far faster to
solve. In practise the unknown time-dependent
exchange-correlation potential must be approximated, introducing 
errors in excitation energies. These are
not always well-understood, although there has been some recent
progress~\cite{AGB03}, for example in the significance of the
asymptotic behaviour of the potential~\cite{CJCS98,TH98b}, and in the
severe frequency-dependence needed for
double-excitations~\cite{MZCB04,CZMB04,C05}).  

It has been found~\cite{DWH03,T03,DH04,GB04b,TAHR99,F01,LLS03,CGGG00,TTYY04}, that excitations involving charge-transfer
(CT) between widely separated species within a molecule are severely
underestimated in TDDFT. This is in contrast to most excitations,
which are accurate to within a few tenths of an electron-volt.  This
has serious consequences. For example~\cite{DH04}, prediction of CT
quenching of fluorescence in light-harvesting bacteria, contrary to
observation.

There have been several recent attempts to overcome this
charge-transfer problem.  In Ref.~\cite{CGGG00}, a $\Delta$SCF
correction is made, while in Ref.~\cite{DWH03}, it is proposed that
TDDFT be mixed with configuration interaction singles. In Ref.~\cite{TTYY04}, a long-range correction using Hartree-Fock is implemented.
Ref.~\cite{T03} suggests a simple improvement by shifting the donor
HOMO and acceptor LUMO energies based on the derivative
discontinuity~\cite{PPLB82}.  Ref.~\cite{GB04b}, on the other hand,  remains purely within TDDFT, and suggests an empirical
asymptotically-corrected kernel that diverges with separation. 

In the present paper, we also stay strictly within TDDFT, and ask, in
contrast to all other approaches, what is the {\it exact} TDDFT
description of charge-transfer in one simple case: charge-transfer
between two open shell species, when the charge transfer states are
well-separated from the other excitations of the system.  By solving a
simple model we deduce what the exchange-correlation kernel must be in
this case.

An important feature of the present analysis involves the step in the
exact ground-state Kohn-Sham (KS) potential that appears between two
widely-separated open shell species of different ionization
potentials. This step is also present in orbital approximations such
as exact exchange (or the KLI approximation to it~\cite{KLI92}). It is
not present in local or semi-local gradient approximations: so our
model for the exact exchange-correlation kernel does not apply to
these cases. Our purpose here is to explore how charge-transfer is
described exactly in a simple model case.

We begin in Section~\ref{sec:TDDFT linear response} with a review of
TDDFT linear response theory, and present the problem of long-range
charge transfer.  A simple model to study this is introduced in
section~\ref{sec:static correlation}. We describe the significance of
the step that develops in the KS potential when a (closed-shell)
molecule composed of two different open-shell species are pulled
apart. The KS ground-state is of a fundamentally different nature than
the true ground-state, resembling the problems of homonuclear
dissociation; the KS charge-transfer energies become zero! We show how
inclusion of the electron-electron interaction breaks the degeneracy
and yields the correct nature of the ground-state and the CT states.
In section~\ref{sec:xc kernel} we describe the implication of static
correlation for the TDDFT exchange-correlation kernel, and derive a
model for it that is valid in the limit that the charge-transfer
excitations are well-separated from all other excitations in the
system. The features are strong-frequency dependence and exponential
dependence on the separation of the two species.  Finally, in
section~\ref{sec:other}, we discuss other ways that static correlation
haunts TDDFT, so far unexplored.

\section{TDDFT Linear Response and CT states} 
\label{sec:TDDFT linear response}
We begin by briefly reviewing the TDDFT linear response formalism. 

Although the density of the KS system is defined to be that of the
interacting system, the excitation energies are not the same: linear
response theory tells us how to correct them.
Applying a small perturbing potential to a ground-state, and measuring
the density response defines the susceptibility, or density-density
response function, $\chi[n_0](\br,\br',t-t') = \delta v\ext(\br
t)/\delta n(\br't')\vert_{n_0}$. The 
susceptibility of the true interacting system is related to that of its non-interacting Kohn-Sham counterpart,
$\chi\s[n_0](\br,\br',t-t') = \delta v\s(\br t)/\delta
n(\br't')\vert_{n_0}$, through an integral equation,
written in the frequency-domain as \cite{PGG96}:
\begin{equation}
\tensor\chi^{-1}(\omega) =\tensor\chi_{\sss S}^{-1}(\omega)-\tensor f_{\sss HXC}(\omega)
\label{eq:fop}
\end{equation}
Here the Hartree-exchange-correlation kernel is the sum 
$f\Hxc[n_0](\br,\br',\omega) = f\H(\br,\br') +
f\xc[n_0](\br,\br',\omega)$. The Hartree kernel is the
density-functional-derivative of the Hartree potential, $f\H(\br,\br')
= 1/\vert \br -\br'\vert$, and the exchange-correlation kernel is that of the exchange correlation potential, $f\xc[n_0](\br,\br', t-t') = \delta v\xc(\br t)/\delta n(\br',t')$.  Transition frequencies of the true system
lie at the poles of $\chi(\br,\br',\omega)$, and oscillator strengths
may be obtained from the residues. The poles of
$\chi\s(\br,\br',\omega)$ are at the KS single excitations; these are
shifted to the true excitations through the action of the
Hartree-exchange-correlation kernel.  So, Eq.~(\ref{eq:fop})
enables us to obtain the interacting excitation energies and
oscillator strengths from the KS susceptibility and the
Hartree-exchange-correlation kernel.  In principle, the exact spectrum
of the interacting system is obtained; in practise, approximations
must be made for (a) the xc contribution to the
ground-state KS potential, and (b) the xc kernel
$f\xc(\omega)$.

In the Lehman representation,
\begin{equation}
\chi({\bf r},{\bf r}';\omega)=\sum_I\left\{\frac{F_I({\bf r})F_I^*({\bf r}')}
 {\omega-\omega_I+i0^+}-\frac{F_I^*({\bf r})F_I({\bf r}')}
 {\omega+\omega_I+i0^+}\right\}, 
\label{eq:chidef}
\end{equation}
where,
$
F_I({\bf r})=\langle 0|\hat{n}({\bf r})|I\rangle
$
with $\hat{n}(\br)$ being the one-body density operator, 
 $I$ labels the excited states of the interacting system, 
and $\omega_I$ is their transition
 frequency. This expression also holds for the KS susceptibility where
 the excited-states are excited Slater determinants and the transition
 frequencies are orbital energy differences.  
Excitations of atoms and molecules are often calculated in a matrix formulation of these equations~\cite{C96}: one lets $q = (i,a)$ be an index representing a single excitation: a
transition from an occupied KS orbital $\phi_i$ to an unoccupied one
$\phi_a$, and  let $\omega_q$ be the difference in the KS
orbital energies, $\omega_q= \epsilon_a - \epsilon_i$. 
Then,   the squares of the true
transition frequencies $\Omega_I =\omega_I^2$
 are the eigenvalues of the matrix
\begin{equation}
\widetilde{\Omega}(\omega)_{qq'}=\delta_{qq'}\omega_q^2+4\sqrt{\omega_q\omega_{q'}}[q|f_{\sss HXC}(\omega)|q'],
\label{eq_casida}
\end{equation}
where 
\ben 
[q|f\Hxc(\omega)|q'] = \int d\br d\br' \phi_i^*(\br)\phi_a(\br) f\Hxc(\br,\br',\omega) \phi_{i'}(\br')\phi_{a'}^*(\br').
\label{eq:sqnotation}
\een
Oscillator strengths of the true system are related to the eigenvectors~\cite{C96}. 

When the coupling between excitations is very small, one may neglect the off-diagonal elements of the matrix, producing the ``small-matrix approximation''~\cite{VOC99,AGB03,GKG97}:
\ben
\omega^2 = \omega_q^2 +4\omega_q[q\vert f\Hxc(\omega_q)\vert q]
\label{eq:sma}
\een
This is equivalent to keeping only the backward and forward
transitions at frequency $\omega_q$ in the Lehman representation of
the response function.
When the shift from the KS transition is small, one may simplify this further to get the ``single-pole approximation''~\cite{PGG96}:
\ben
\omega = \omega_q + 2[q|f\Hxc(\omega_q)|q]\,.
\label{eq:spa}
\een

Now, consider applying the theory to long-range charge-transfer excited states. In the limit of
 large separations $R$, the {\it exact} energy cost for transferring an electron from
donor to acceptor, is
\ben
\omega = I_D - A_A - 1/R
\label{eq:CTenergy}
\een
where $I_D$ is the ionization energy of the donor, and $A_A$ is the
electron affinity of the acceptor.  As discussed in Refs.~\cite{DWH03,T03},
the failure of TDDFT to reproduce this is evident from a single-pole analysis (Eq.~(\ref{eq:spa})), where
\bea
\nonumber
\omega &=& \epsilon_A^L - \epsilon_D^H + \int d^3r_1d^3r_2 F(\br_1)f\Hxc(\br, \br',\omega_{q}) F(\br_2)\\
&\approx&\epsilon_A^L - \epsilon_D^H = I_D - A_{\sss S, A}
\eea
Here $F(\br) = \phi_D^H(\br)\phi_A^L(\br)$ is the product of the HOMO of the donor (KS orbital energy $\epsilon_D^H$) and the LUMO of the acceptor (KS orbital energy $\epsilon_A^L$). 
Because there is exponentially small overlap between the atomic
orbitals on the widely separated donor and acceptor, the integral term vanishes, and
the TDDFT energy collapses to the bare KS energy, as indicated in the second line. We have used the
fact that the KS HOMO energy is exactly the negative of the ionization energy,
 but the LUMO energy differs from the negative affinity by the discontinuity~\cite{P85b,PPLB82,PL97c,AB85}:
\ben
A = A\s + A\xc  = -\epsilon_L + A\xc
\een 
  Now, common
approximations underestimate $I$, but how does {\it exact} TDDFT get
the exact energy? That is, suppose we had the exact ground-state
Kohn-Sham potential and also the exact exchange-correlation
kernel. Then the exact TDDFT HOMO energy is indeed $I_D$, but
 how does the exact TDDFT retrieve the discontinuity $A\xc$, and the $-1/R$ Coulomb fall-off? We address this in the next two sections.  

\section{CT between two open shell species: the role of static correlation}
\label{sec:static correlation}
\begin{figure}
 \centering 
 \includegraphics[height=4cm,width=6.5cm]{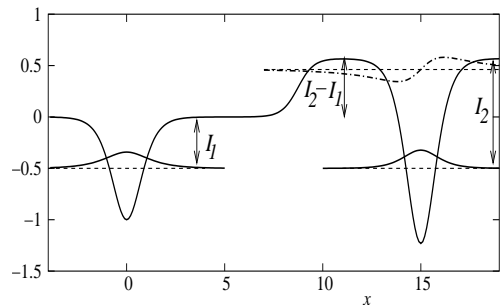} 
\caption{The 2-electron KS potential for two widely separated ``Eckart'' atoms, $V\ext(x) = 1/\cosh^2(x) +1.8/\cosh^2(x-15)$. The highest occupied orbitals of each atom are shown solid, and the first excited state of the right-hand atom is shown in dash-dot.} 
\label{fig:eckart}
\end{figure}

To study this question, we consider first the simplest model: electron
transfer between two one-electron neutral ``atoms'', separated by a large
distance $R$. Let $I_{1(2)}$ be the ionization energy of atom 1(2); then 
the atomic orbital occupied in the ground-state has energy $\epsilon_H = -I_{1(2)}$, where the $H$
subscript stands  for HOMO of the atom.
When we consider the closed-shell molecule composed of the two atoms at large separation,
 the
ground-state KS potential develops a step between the atoms, that
exactly lines up the atomic HOMOs~\cite{P85b,AB85b}. To see this, consider
the Kohn-Sham wavefunction $\Phi_0$ for the molecule. This is the doubly-occupied bonding orbital:
\bea
\nonumber
\Phi_0 &=& \phi_0(\br_1)\phi_0(\br_2)\left(\vert\uparrow \downarrow\rangle  - \vert\downarrow\uparrow\rangle\right)/\sqrt{2}\,, \; {\rm with}\\
\phi_0(\br) &=& \left(\phi_1(\br) +\phi_2(\br)\right)/\sqrt{2(1+S_{12})}
\label{eq:bo}
\eea
where $S_{12}$ is the overlap integral $\int\phi_1(\br)\phi_2(\br) d^3r$, exponentially small in the separation $R$. 
In the limit of large separation, $\phi_{1,2}$ denotes the occupied 
atomic orbital of atom 1(2). 
Inserting $\phi_0$ into the KS equation, while recognizing that near
each atom, the orbital must reduce to that atom's atomic HOMO,
with the appropriate orbital energy, one sees that the
exchange-correlation potential must develop a step of size $\vert I_2
- I_1\vert$ between the atoms; raising the potential of the atom with
the larger ionization energy.  A simple 1-d example is given in Fig.~\ref{fig:eckart} where each ``atomic nucleus'' is an Eckart potential. The potential eventually goes back down to zero on the right-hand-side. (A similar picture for two delta-function ``nuclei'' can be found in Ref.~\cite{P85b}). 
The step renders the HOMOs of each atom
degenerate, and is necessary to prevent dissociation into fractionally
charged species~\cite{PPLB82}. If they were not exactly lined up, then one 
could
lower the energy of the system by transferring a fraction of charge
from one of the atoms to the other, contradicting the charge neutrality that
is observed in nature. This phenomenon is closely related to the derivative discontinuity discovered in the 80's~\cite{PPLB82,PL97c,P85b,AB85,AB85b}. 
It is important to note that the step is present in exact DFT and is approximated in spatially-nonlocal approximations such as exact-exchange~\cite{GKKG98,KLI92}
Most common approximate potentials, such as LDA, garner only local or semi-local density information, and so do not display the step. 

The fact that in exact DFT the atomic HOMO's become degenerate has an
apparently drastic consequence for CT excited states:
Whereas a (semi-)local DFT approximation would yield a
finite KS transition energy, albeit underestimated, the {\it exact} KS system yields
{\it zero} energy for charge transfer between the atoms!

The Kohn-Sham energies are poor because, in this case, the KS
description is fundamentally different from the interacting
system. The KS ground-state involves each electron delocalized over
both atoms, whereas the true wavefunction has one atom on
each. We shall come back to this point shortly. 
 
The alignment of the atomic HOMO's leads to three nearly degenerate
singlet two-electron KS states, whose spatial parts are:
\bea
\nonumber
\Phi_0& =& \phi_0(\br_1)\phi_0(\br_2)\\
\nonumber
\Phi_q &=& \left(\phi_0(\br_1)\bar{\phi_0}(\br_2) + \bar{\phi_0}(\br_1)\phi_0(\br_2)\right)/\sqrt{2}\\
\Phi_D& =& \bar{\phi_0}(\br_1)\bar{\phi_0}(\br_2)
\label{eq:3by3}
\eea
where $\phi_0$ is the bonding orbital of Eq.~(\ref{eq:bo}), and $\bar{\phi_0}$
is the antibonding orbital
\ben
\bar{\phi_0}  = \left(\phi_1 - \phi_2\right)/\sqrt{2(1-S_{12})} \,.
\een
At large separations the ``singly-excited'' $\Phi_q$ is slightly
higher in energy than $\Phi_0$ by tunnel splittings between the atomic
orbitals, and the ``doubly-excited'' $\Phi_D$ higher again; the orbital energy of the antibonding state is at $\bar\omega \sim \exp(-R)$, where $R$ is the separation between the atoms. The proximity of $\Phi_q$ and $\Phi_D$ to $\Phi_0$ is a signature of the static correlation in the ground-state. 

Before discussing the implication of this static correlation for TDDFT, we
 first show that the electron interaction splits the degeneracy
 between the Kohn-Sham states, recovering states of CT
 nature and approximations for their energies.  We assume that the
 tunnel-splittings are much smaller than the energy-splitting from the
 electron-electron interaction, and that the
 basis of the three states, Eqs.~(\ref{eq:3by3}), has negligible couplings to all other KS
 excitations. To simplify the discussion, we choose $I_2 > I_1$. 
Diagonalizing the true Hamiltonian in this basis, yields the three states:
\newline
(i) the Heitler-London ground-state
$\Psi_0 = \left(\phi_1(\br)\phi_2(\br') +   \phi_2(\br)\phi_1(\br')\right)\sqrt{2}$
\newline
(ii) charge-transfer excited state  from atom 2 to atom 1: 
$\Psi_{2\to 1} = \phi_1(\br)\phi_1(\br')$. 
This has energy, relative to the Heitler-London state, 
\ben
\omega_1 = I_2 - A_1 - 1/R
\label{eq:om1}
\een
where $A = A\s +A\xc^{\sss approx}$, with
\ben
 A\xc^{\sss approx} = -\int d^3r\int d^3r' \phi_H(\br)^2\phi_H(\br')^2 V\ee(\br - \br')
\label{eq:Axcapprox}
\een
Here $V\ee$ is the electron-electron interaction,  $V\ee = 1/\vert \br - \br'\vert$.
For later purposes, the subscript $H$ stands for the highest occupied orbital of the atom; for one-electron atoms, it simply indicates the occupied atomic orbital on the atom.
Note also that for open-shell atoms, $A\s = I =-\epsilon_H$.
\newline
(iii) charge-transfer excited state from atom 1 to atom 2:
$\Psi_{1 \to 2} = \phi_2(\br)\phi_2(\br')$, of energy
\ben
\omega_2 = I_1 - A_2 - 1/R
\label{eq:om2}
\een

Thus, the correct nature of the ground-state, and CT
states with the correct form of energy (c.f. Eq.~\ref{eq:CTenergy}), including
an approximation for the exchange-correlation contribution to the
electron affinity, are recovered. 

In a sense, accounting for the electron-interaction un-does the static
correlation in the system.  Because of the step in the potential, and
the consequent nearly degenerate triple of KS determinants, the
ground-state problem resembles  the well-known homonuclear
dissociation problems of ground-state DFT~\cite{PSB95,GGGB00}, where the
dissociation limit is incorrect because the KS ground-state involves
mixing ionic combinations into the true neutral Heitler-London form.
Static correlation was an
important agent above in treating the perturbation: The Hamiltonian can always be written in terms of the
KS Hamiltonian $H\s$, as $H = H\s +V\ee - v\H -v\xc$. For any
two-electron system, as we have here, $v\x = -v\H/2 \sim O(\lambda)$,
where $\lambda$ is the interaction strength.  Typically the
correlation potential is $O(\lambda^2)$, appearing at the next order
of interaction strength, but when there is static correlation, it is
much stronger: $v\c$ cancels $v\H +v\x$ and adds the step, yielding
\bea
\nonumber
v\Hxc(\br) &=& 0\,,\,\mbox{for}\; \br\; \mbox{near atom 1}\\
\nonumber&=&I_2 -I_1\,,\,  \mbox{for}\; \br\; \mbox{near atom 2}
\eea
This must be so in order for the KS equation for the molecular orbital
$\phi_0$ to reduce to that for the atomic orbital $\phi_{1,2}$,
respectively, near atom 1 and atom 2.  We discuss how static
correlation affects the exchange-correlation kernel of TDDFT in the next section.

\begin{figure}
 \centering 
 \includegraphics[height=4cm]{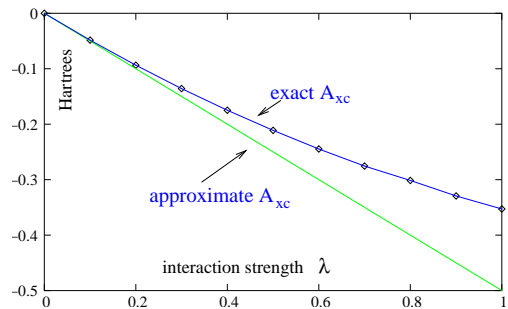} 
\caption{Exchange-correlation contribution to the electron affinity of a fermion in a $\delta$-well:
exact (extracted from Ref.~\cite{R71}) and approximate
(Eq.~\ref{eq:Axcapprox})}
\label{fig:example}
\end{figure}

We now consider the approximate result for $A\xc$,
Eq.~(\ref{eq:Axcapprox}). The derivation neglected all KS states
except for the three in the basis~(\ref{eq:3by3}): the approximation
thus becomes exact when the basis is truly isolated, or in the limit of
weak interaction so that the higher-lying KS states
are not appreciably mixed in.
 A simple example for
which the exact affinity at all interaction strengths is easily calculable, is that of a fermion in
one dimension living in a delta-function well (i.e. a one-dimensional H atom).  It is a standard
textbook problem to find the only bound-state of the well.
The  electron affinity can be obtained by subtracting
its ground-state energy from that of two interacting fermions in the
well.  When their interaction is a delta-function repulsion of
strength $\lambda$:
\ben
H = -\sum_i^2 \frac{1}{2}\frac{d^2}{dx_i^2}- \sum_i^2\delta(x_i) + \lambda\delta(x_1 -x_2)\,,
\een
the ground-state energy has an exact solution~\cite{R71}.  (Atomic units are used throughout the paper). In
figure~\ref{fig:example}, the exact $A\xc$ has been plotted from $A\xc
= E(1) - E(2) -A\s = -E(2) - 2I$, where the ground-state energy for two
fermions, $E(2)$, is obtained from Ref.~\cite{R71}, and that for one
fermion is $E(1) = -A\s = -I = -0.5H$.  Comparing this with the
approximation from Eq.~(\ref{eq:Axcapprox}), where $\phi_H$ is the
textbook solution for the orbital in the one-fermion problem, we see that, 
as expected, the approximation becomes exact in the weak interaction limit. 

This result for $A\xc$ is consistent with a perturbative analysis of
$E(N) - E(N+1)$. Consider adding one electron to an $N=1$-species. In
an unrelaxed approximation, the extra electron plops into the same
spatial orbital as the first. A perturbative approximation for the
electron affinity is obtained from calculating the difference in the
Hamiltonians of the 2-electron and 1-electron system, in the unrelaxed
doubly-occupied orbital. One finds exactly the
result, Eq.~(\ref{eq:Axcapprox}).

Although our results so far have been derived for charge transfer
between two one-electron species, it is straightforward to show that they
 hold for charge-transfer
between two general open-shell species. There the KS molecular HOMO
sits atop a spin-paired KS determinant, containing all the inner
orbitals of both atoms.  

So far, we have examined the exact ground-state KS potential for a
closed shell molecule composed of two open-shell species, and found
that the step in between the species leads to a near zero value for
charge-transfer in the Kohn-Sham system. We showed that the static
correlation is broken by the electron-electron interaction, and how the latter allows us to
recapture the true ground-state and excited charge-transfer states,
together with an approximation for the exchange-correlation electron
affinity.

\section{The Exchange-correlation kernel for charge-transfer}
\label{sec:xc kernel}
How does  the static correlation in the ground-state affect the TDDFT description of charge-transfer? In TDDFT, the exchange-correlation kernel must play the role of the diagonalization of the previous section in salvaging the
CT energies from their near zero KS value. 
We now examine
what the nature of the exact $f\xc$ must be, and the signatures of static
correlation in the kernel. 

Consider the KS response function, $\chi\s(\br,\br',\omega)$. 
We may write this as
\ben
  \chi\s(\br,\br',\omega) \approx \frac{2\bar{\omega}}{\omega^2 -(\bar\omega)^2}X\s(\br,\br',(\omega))
\label{eq:chis}
\een
where we include only the contribution to Eq.~(\ref{eq:chidef}) from the
antibonding transition, valid in the limit that the KS bonding-antibonding pair is well-separated from the higher KS transitions. 
 The assumption is that to lowest order in the 
interaction strength, the interacting charge-transfer states arise from 
this transition only.
This is analogous to the approximation of restricting to the
$3\times3$ subspace when we considered diagonalization.
The numerator $X\s(\br, \br',(\omega)) = \left(\phi_1^2(\br) - \phi_2^2(\br)\right)\left(\phi_1^2(\br') - \phi_2^2(\br')\right)/4 + O\left((\omega^2 - \bar\omega^2)\Delta\s/[\bar\omega(\omega^2 - \Delta\s^2)]\right)$, where $\Delta\s$ is the frequency of the next highest KS transition. 

Notice that $\chi\s$ vanishes exponentially with separation of the
atoms due to the exponential dependence of the tunnel splitting
$\bar\omega$ on separation. This can be understood from $\chi\s =
\frac{\delta n}{\delta v\s}(\omega)$: the perturbation at $\bar\omega$
excites the antibonding transition, whose density difference from the
bonding orbital's is exponentially small with separation.

Now we turn to the interacting response function, $\chi(\br, \br',
\omega)$, which is a sum over all excitations of the interacting
system. As in the KS case, we zoom into the excitations that describe
charge transfer:
\ben
\chi(\br,\br',\omega) \approx\frac{2\omega_1}{\omega^2 -\omega^2_1}X_1(\br,\br',(\omega)) + \frac{2\omega_2}{\omega^2 -\omega^2_2}X_2(\br,\br',(\omega))
\label{eq:chi}
\een
Within the isolated basis approximation, $X_1(\br, \br',(\omega)) =X_2(\br, \br',(\omega)) = 2\phi_1(\br)\phi_2(\br)\phi_1(\br')\phi_2(\br') + O\left(1/\Delta\right)$. 
Again the response function vanishes exponentially with separation,
but this time because the CT transitions have
exponentially weak oscillator strength due to overlap between
widely separated spatial regions: in the orbital product $\phi_1(\br)\phi_2(\br)$, each orbital is exponentially small as a function of atomic separation $R$ in the places where the other is finite.
The inverse functions, $\chi^{-1}$ and $\chi\s^{-1}$, are therefore both exponentially large as a function of separation $R$ ({\it not} to be confused with the finite behaviour as a function of $\vert \br - \br'\vert$). 

We note that we are retaining backward transitions in the expressions
above: these are ``small-matrix approximations'' rather than
``single-pole approximations'' that include only the forward
transition. The latter is not valid here because the antibonding
transition frequency is very small: the backward transition is of
almost the same magnitude as the forward, so should not be neglected.
The single-pole approximation is
only valid when the shift from the KS energy is small~\cite{AGB03}:
here, the shift is the entire CT energy, since the KS transition
energy is near zero (i.e. exponentially small in the separation).

As we saw in the diagonalization procedure, the  CT
transitions of the interacting system arise from the KS subspace of bonding and antibonding
orbitals. In the diagonalization procedure, the number of states was
preserved: the three KS 2-electron singlet molecular determinants of Eq.~(\ref{eq:3by3}),
composed of the bonding and anti-bonding orbitals, were rotated by the
electron-electron interaction into the Heitler-London ground-state and
CT states onto each of the two atoms. The picture is
somewhat different in the response functions. Here the space is of
excitations out of the ground-state, so in the interacting case there
are two (one to each CT state). But the KS response function
has only the single-excitation (corresponding to $\Phi_0 \to \Phi_q$
of Eq.~(\ref{eq:3by3})): double excitations cannot appear in the KS
response function because the numerator involves the one-body density
operator, which connects only states differing in one orbital (see
also Ref.~\cite{MZCB04}). It is the job of the exchange-correlation kernel, 
$f\xc$ to generate an extra pole, and also to mix them. Specifically, 
\bea
\nonumber
f\Hxc(\omega) = \chi\s^{-1}(\omega) - \chi^{-1}(\omega)&&\\
=\frac{1}{2}\left(\frac{(\omega^2 - \bar\omega^2)}{\bar\omega}X\s^{-1} - \frac{(\omega^2 - 
\omega_1^2)(\omega^2 -\omega_2^2)}{\omega_2(\omega^2 - \omega_1^2)+\omega_1(\omega^2 - \omega_2^2)}X_1^{-1}\right)&&
\label{eq:fhxc1}
\eea
Due to the $f$-sum rule, in the limit of the isolated basis, 
\ben
\bar\omega X\s = \omega_1 X_1 +\omega_2 X_2 \approx (\omega_1 + \omega_2)X_1
\een
we find $X_1^{-1} = (\omega_1 + \omega_2)X\s^{-1}/\bar\omega$. 
Finally, we require that a dressed small-matrix approximation, arising
from plugging Eq.~(\ref{eq:fhxc1}) into Eq.~(\ref{eq:sma}) yields the true transition frequencies $\omega = \omega_1$ and $\omega = \omega_2$ as solutions. This fixes $[q\vert X\s^{-1} \vert q] = 1/2$, and our expression for the kernel becomes:
\ben
\bar\omega[q\vert f\Hxc(\omega)\vert q] = \omega_{av}^2 + \frac{\omega_1\omega_2 -\bar{\omega}^2}{4} + \frac{\omega_1\omega_2\omega_{av}^2}{\omega^2 - \omega_1\omega_2}
\label{eq:result}
\een 
for the KS transition $q=$bonding$\to$anti-bonding orbital, and where $\omega_{av} = (\omega_1 +\omega_2)/2$.
 This, with $\omega_1$ and $\omega_2$ given by Eqs.~(\ref{eq:om1}), (\ref{eq:Axcapprox}),and (\ref{eq:om2}), gives the
exact exchange-correlation kernel matrix element in terms of KS quantities, in the limit that the
charge-transfer states are isolated from the other transitions, in
that coupling to them can be neglected.
Notice the strong non-adiabaticity in the exact kernel, manifest by
the pole in the denominator on the right-hand-side.

In the usual presentations of the TDDFT charge-transfer problem, the
 failure is due to the vanishing overlap between the occupied orbital,
 on one atom, and the unoccupied one on the other atom, widely
 separated. This means that $q = \phi_i\phi_a$ vanishes exponentially
 with separation.  This does not occur in the exact analysis of CT
 between open shells presented here: here the occupied orbital is the
 bonding orbital and the unoccupied is the anti-bonding orbital, so
 their overlap goes as $(\phi_1^2 - \phi_2^2)/2$ where $\phi_1$ is the
 atom 1's HOMO (or LUMO) and $\phi_2$ is atom 2's HOMO (or LUMO).
 Although the overlap remains finite, the matrix element $[q\vert
 f\Hxc \vert q]$ diverges exponentially with interatomic separation
 $R$, as $1/{\bar{\omega}}$, as can be seen from
 Eq.~(\ref{eq:result}).

A divergence of $f\Hxc$-matrix elements has previously been found in
  H$_2$ dissociation~\cite{GGGB00}, where the the lowest
  singlet-singlet KS transition energy vanishes with separation, while
  the true energy approaches a finite number;
the step in the KS potential for our heteroatomic molecules leads to a
similar feature raising its head again here, but now with the kernel
being strongly frequency-dependent. Also, the CT treatment in
Ref.~\cite{GB04b} utilized an empirically determined kernel that
displays also exponentially large behaviour as a function of atomic
separation.




\section{Discussion and Other implications of Static Correlation}
\label{sec:other}
By considering electron transfer between two one-electron species at
long-range, we investigated how charge-transfer is captured in the
exact KS system. The bare KS energies for CT approach zero
exponentially with the separation of the species, but including the
electron-electron interaction, splits the near-degeneracy and
recaptures finite CT energies of the correct form.  We derived the
form of the TDDFT exchange-correlation kernel, Eq.~(\ref{eq:result}), that
becomes exact in the limit that the charge-transfer states are
isolated from all the other excitations in the system. The main
features of the kernel are (i) a strong frequency-dependence, due to
the mixing in of the double excitation to the antibonding KS state,
and (ii) a dependence on the inverse tunnel splitting between the two
species, that goes exponentially with their separation. 

A crucial feature of the exact KS system in this paper is the step
 between the two widely separated species. This gives rise to static
 correlation in the ground-state, and hence the strong
 frequency-dependence and exponential-dependence on the atomic
 separation of the $f\xc$ matrix element, discussed above. Not only
 does it play a vital role in the description of CT states, but it
 also has very interesting, and unexplored, effects on other
 excitations of a long-range heteroatomic molecule.  For example,
 higher atomic excitations become KS resonances, as illustrated in
 Figure~\ref{fig:eckart}. Here two one-dimensional ``Eckart atoms''
 are shown: the second excited state of the atom on the right can
 tunnel out of the barrier presented by the step. The acrobatics the
 exchange-correlation kernel must perform in order to turn such a
 resonance back into a bound state of the true system, will be pursued
 in future work.

Another interesting consequence of the step is that for {\it every}
single excitation out of the HOMO bonding orbital, $\phi_0\to \phi_a$, there is
a nearly degenerate double-excitation, $(\phi_0,\phi_0) \to (\bar\phi, \phi_a)$, where
the other electron occupying the HOMO is excited to the antibonding
orbital, separated in energy only by the tunnel-splitting. Doubles are
absent in the KS response~\cite{MZCB04}, but they are essential in
this case for a correct prediction of the nature and energy of the
interacting state: for example, it is needed to lead to a
charge-neutral excited state on each atom~\cite{unpub}. The presence of double
excitations immediately leads to poles in the frequency-dependence of
the exchange-correlation kernel~\cite{MZCB04}; the static correlation 
implies the poles are ubiquitous and will be investigated in future work.

{\it Acknowledgement} This work is supported in part by the American Chemical Society's Petroleum Research Fund and the Research Foundation of CUNY.

\end{document}

%% file: dft.tex
\def\bea{\begin{eqnarray}}
\def\eea{\end{eqnarray}}
\def\ben{\begin{equation}}
\def\een{\end{equation}}
\def\benu{\begin{enumerate}}
\def\enu{\end{enumerate}}

\def\sss{\scriptscriptstyle\rm}

\def\1var{(\bx_1...\bx\N)}

\def\br{{\bf r}}

\def\bx{{x}}

\def\x{_{\sss X}}
\def\c{_{\sss C}}
\def\s{_{\sss S}}
\def\xc{_{\sss XC}}
\def\Hxc{_{\sss HXC}}

\def\N{_{\sss N}}
\def\H{_{\sss H}}

\def\ext{_{\rm ext}}

\def\ee{_{\rm ee}}

\def\sph_int{ {\int d^3 r}}
